\documentclass[11pt]{article} 
\usepackage{amsmath} 
\usepackage{amsthm} 
\usepackage{amssymb}	
\usepackage{graphicx} 
\usepackage{multicol} 
\usepackage[dvips,letterpaper,margin=0.75in,bottom=0.5in]{geometry}
\makeatletter 
\renewcommand{\maketitle} 
{ \begingroup \vskip 10pt \begin{center} \large {\bf \@title}
	\vskip 10pt \large \@author \hskip 20pt \@date \end{center}
  \vskip 10pt \endgroup \setcounter{footnote}{0} }
\makeatother 
 
\newcommand{\abs}[1]{\left| #1 \right|} 
 
\newcommand{\ket}[1]{\left| #1 \right>} 
\newcommand{\bra}[1]{\left< #1 \right|} 
\newcommand{\braket}[2]{\left< #1 \vphantom{#2} \right|
 \left. #2 \vphantom{#1} \right>} 
\newcommand{\matrixel}[3]{\left< #1 \vphantom{#2#3} \right|
 #2 \left| #3 \vphantom{#1#2} \right>} 
\let\baraccent=\= 
\renewcommand{\=}[1]{\stackrel{#1}{=}} 

\theoremstyle{definition}

\theoremstyle{remark}

\usepackage{hyperref}
\usepackage{amsmath}
\usepackage{amssymb}
\renewcommand{\title}[1]{%
    \bigskip%
    \begin{center}%
    \Large\bf #1%
    \end{center}%
    \vskip .2in}

\renewcommand{\author}[1]{%
    {\begin{center}
    #1
    \end{center}}}
\newcommand{\address}[1]{\vspace{-1.7em}\vspace{0pt}
    {\begin{center}
    \it #1
    \end{center}}}

\begin{document}

\title{ $PT$ symmetry and a dynamical realization of the SU(1,1) algebra}

\author{Rabin Banerjee},
 
\address{S. N. Bose National Centre for Basic Sciences \\
JD Block, Sector III, Salt Lake City, Calcutta -700 098, India.}
 
\author{ Pradip Mukherjee}
 
\address{Barasat Government College\\
10 KNC Road, Barasat, Kolkata -700 124, India.
}

\begin{abstract}
{ We show that the elementary modes of the planar harmonic oscillator can be quantised in the framework of quantum mechanics based on pseudo-hermitian hamiltonians. These quantised modes are demonstrated to  act as
dynamical structures behind a new Jordan - Schwinger realization of the
SU(1,1) algebra. This analysis complements the conventional Jordan - Schwinger construction of the SU(2) algebra based on hermitian hamiltonians of a doublet of oscillators.
 }
\end{abstract}
Traditionally quantum mechanics was formulated with hermitian hamiltonians which ensures real eigenvalues, unitarity etc which are the expected attributes of a physically meaningful theory.
The observation that nonhermitian operators may also have real eigenvalues was not considered
very important in quantum theory. The situation has been drastically altered over the last decade or so. Now we definitely know that consistent quantisation is possible with nonhermitian hamiltonians if the theory is $PT$ symmetric \cite{B}. The corresponding hamiltonians satisfy $\eta^{-1} H^\dagger \eta = H$ where $ \eta = PT$\footnote{However, there exist more general types of complex hamiltonians which may have real eigenvalues}. Such operators will henceforth be called $\eta$ hermitian,
 $PT$ symmetric theories are not only of theoretical interest now a days; such theories have been identified in experimental systems in diverse fields e.g
in superconductivity, quantum optics etc \cite {ap4}. Naturally, study of $PT$ symmetric theories has occupied a large portion of the recent literature. But simple systems which have been quantised long back in the usual way {\it{ may have $PT$ symmetry}}. Then alternative quantisation of such systems may be undertaken. This may reveal some more interesting physics. In this paper we take the bidimensional oscillator as an example and show that the quantisation of the system based on $PT$ symmetry can be exploited to construct a new Jordan - Schwinger (J - S) map \cite{js1,js2} of the SU(1,1) algebra.

The harmonic oscilator problem has been ubiquitious in the realm of physics. The quantum mechanical
 theory of the harmonic oscillator is exact and is the starting point of analysis in many fields. Out of
the myriads of application of the quantum harmonic oscillator a beautiful piece is  the Jordan - Schwinger 
(J - S) construction \cite{js1,js2} of the SU(2) algebra using a doublet of oscillator algebras \cite{SK}. The generators give rise to a Casimir operator which is factorisable in the usual way as should happen for the angular momentum algebra. There is no surprise here because the doublet of oscillator algebras has a dynamical structure behind it, namely the two dimensional oscillator. The total angular momentum quantum number $J$ may be obtained from two elementary rotor's angular momenta by the well known angular momentum addition rules. However, the question of rotation symmetry does not arise for the usual one dimensional harmonic oscillator. What then are the one dimensional elementary modes that should correspond to the individual oscillator algebra? 

 In the recent past it has been shown that a planar oscillator can be synthesised from two oscillator lagrangians of opposite chirality that carry  opposite charges corresponding to the SU(2) symmetry group \cite{BG}. These individual pieces have been termed chiral oscillators. It is now pretty obvious
 that the chiral oscillators constitute the elementary structures of the
 dynamical model behind the J - S construction of the SU(2) algebra. Indeed it is
observed \cite{BG} that corresponding to one quantum of excitation of the left
handed chiral oscillator (energy $\omega$) 
the angular momentum projection is $\frac{1}{2}$
whereas that for the right handed one it is $-\frac{1}{2}$.
Such an identification is important because a complete dynamical realization of the symmetry is thus available. In this connection one recalls that the same doublet of oscillator algebras was used to construct J--S type realisation of the SU(1,1) and other polynomial algebras \cite{SBJ, ap1}. However these constructions are at the algebraic level. No dynamical realization was given . In fact the Casimir operator constructed for SU(1,1) in \cite{SBJ} is not factorisable, contrary to what happens for SU(2) \cite {SK}, indicating that perhaps there could be no proper dynamical realisation. 

  The J--S type construction which will be presented here is fundamentally new. It exploits a different lagrangian model of the bidimendional oscillator and is based on a novel quantum mechanics \cite{BM} with pseudo hermitian hamiltonians \cite{B}. The genesis of this new lagrangian model lies in the not so widely known fact that
apart from the usual (direct) representation, there is a different (indirect) lagrangian representation of  the planar harmonic oscillator \cite{S}. This Lagrangian has SU(1,1) symmetry as will be shown subsequently. The SU(1,1) symmetry of the bidimensional oscillator is not much emphasised in the literature. The question naturally poses itself -- is it possible to synthesise the bidimensional oscillator lagrangian from elementary oscillators having opposite 'chirality' with respect to SU(1,1) symmetry? If it is so, can one have a dynamical realisation of the J - S type construction of the SU(1,1) algebra, exploiting the (indirect) oscillator representation? We show in the present paper that this can be done in the fold of $PT$ symmetric \cite{B} quantization \cite{BM} of the harmonic oscillator. Note that the J--S map is of wide use in different branches of physics \cite{ap1}, specifically in quantum information theory \cite{ap2} and quantum optics \cite{ap3}. On the other hand $PT$ symmetry has found applications in many of the fields \cite{ap4}. Thus the problem addressed here has much interest in view of the current research.  

The indirect lagrangian formulation of the bidimensional oscillator is characterised by the fact that here the equations of motion of one oscillator is obtained by varying the action wth respect to the other coordinate and vice versa. Thus
to find the Lagrangian of the planar harmonic oscillator in the indirect approach we consider the oscillators \footnote{Both $x$ and $y$ are real}
\begin{eqnarray}
\ddot{x}  + \omega^2 x = 0\label{1}\\
\ddot{y} + \omega^2 y = 0\label{6}
\end{eqnarray}
and write the variation of the action as
\begin{equation}
\delta S = \int^{t_2}_{t_1}dt\left[\left(\frac{d}{dt}\dot{x} 
               + \omega^2 x\right)\delta y +
                \left(\frac{d}{dt}\dot{y} + \omega^2 y\right)\delta x\right]
                          \label{7}
\end{equation}
From (\ref{7}), equation (\ref{1}) is obtained by varying $S$ with $y$
whereas (\ref{6}) follows from varying $S$ with $x$.
Since the equations of motion for $x$ and $y$ follow as Euler - Lagrange 
equations for $y$ and $x$ respectively, the method is called the indirect
method. Now, starting from (\ref{7}) we can deduce
\begin{equation}
\delta S = -\delta \int^{t_2}_{t_1}dt\left[\dot{x}\dot{y} -\omega^2 x  y\right]
                          \label{8}
\end{equation}
It is then possible to identify
\begin{equation}
L_I = \dot{x}\dot{y} -\omega^2 x  y
                          \label{9}
\end{equation}
as the appropriate Lagrangian in the indirect representation
. The equations of motion following from this lagrangian are just equations (\ref{1}) and (\ref{6}).The direct lagrangian on the other hand, has the mundane structure
\begin{equation}
L_{D} = \frac{1}{2}\left(\dot{x}^2 - \omega^2x^2\right) + \frac{1}{2}\left(\dot{y}^2 - \omega^2y^2\right)
\label{1111}
\end{equation}
where, varying with $x$ (or $y$) yields (\ref{1}) (or (\ref{6})).

 The Lagrangian (\ref{9}) can be written in a suggestive way, mimicking (\ref{1111}) by the 
substitution of the hyperbolic coordinates $x_1$ and $x_2$
\cite{BGPV} defined by 
\begin{eqnarray}
x &=& \frac{1}{\sqrt{2}}(x_1 + x_2)\nonumber\\
y &=& \frac{1}{\sqrt{2}}(x_1 - x_2)\label{10} 
\end{eqnarray}
We find that the Lagrangian $L_{I}$ becomes
\begin{equation}
L_{I} = \frac{1}{2}\dot{x}_1^2 - \frac{\omega^2}{2}x_1^2
- \frac{1}{2}\dot{x}_2^2 + \frac{\omega^2}{2}x_2^2\label{111}
\end{equation}
The above Lagrangian can be expressed in a notationally elegant form \cite{BM}
\begin{equation}
L_I = \frac{1}{2}g_{ij}\dot{x}_i\dot{x}_j - \frac{\omega^2}{2}g_{ij}x_i x_j
                  \label{11}
\end{equation}
by introducing the pseudo - Eucledian metric $g_{ij}$ 
given by $g_{11} = -g_{22}
= 1$ and $g_{12}$ = 0.
{\vspace {0.7cm}}

Note that the composite Lagrangian (\ref{11}) 
is invariant under the transformation
\begin{equation}
x_i \to x_i + \theta\sigma_{ij}x_j\label{24}
\end{equation}
where $\sigma$ is the first Pauli matrix. The corresponding symmetry group is easily recognised as SU(1,1). Thus equation (\ref{24}) represents a  SU(1,1) rotation in the plane. Apart from the continuous symmetry (\ref{24}), the theory has the discrete symmetry 
\begin{equation}
x_i \to x_i^\prime = g_{ij}x_j\label{24d}
\end{equation}
The origin of (\ref{24d}) is the combined action of parity $P$ and time reversal $T$. This can be seen as follows. Under $P$, $x$ and $y$ transform as \cite{ap4},
\begin{equation}
x \to y \hskip .3cm \rm{and} \hskip .3cm y \to x\label{P}
\end{equation}
while under $T$
\begin{equation}
x \to x \hskip .3cm \rm{and} \hskip .3cm y \to y\label{T}
\end{equation}
 Using these in equation (\ref{10}) we arrive at (\ref{24d}).
Thus the theory
(\ref{11}) is $PT$ symmetric \cite{B}. 

 Our primary aim is to find the elementary modes of the bidimensional oscillator (\ref{11}). This lagrangian reduction can be done using the soldering
formalism 
which has found applications in various contexts 
\cite{BW},
 \cite{BK2}, \cite{BK1},
 \cite{A}. The elementary modes are 
\begin{equation}
L_{\pm} = 
\pm i\omega\epsilon_{ij}x_i\dot{x}_j - \omega^2 g_{ij}x_i x_j\label{23}
\end{equation}
Before we discuss the proof of the statement 
note that there is a factor of $i$ in the Lagrangians (\ref{23}). This makes the Lagrangians complex. The corresonding hamiltonian is also complex. However,  a great success of theoretical efforts over the last decade is to discover a formalism  \cite{B} to quantize theories with complex hamiltonians \cite{B}, A sufficient condition is found to be the existence of $PT$
symmetry. Since we have seen that our theory has this symmetry (see equation (\ref{24d})), it is quantisable. Use will be made of a novel quantisation method developed, exploiting the  $PT$
symmetry, in \cite{BM}. There the complex hamiltonians satisfied
the condition $\eta^{-1}H^{\dagger}\eta = H $ with  $\eta = PT$. We will name such operators as $\eta$ hermitian in the following.
   
  Let us now discuss the soldering of (\ref{23}) leading to (\ref{11}). We start from a simple sum
\begin{equation}
L(y,z) = L_+(y) + L_-(z)\label{s1}
\end{equation}
 The  soldering  now proceeds as follows.
 Use $x_i = y_i - z_i$ in $L(y,z)$ to eliminate
$z_i$ so that
\begin{eqnarray}
L(y,x) = & &i2\omega\epsilon_{ij}\left(y_i \dot{x}_j -\frac{1}{2} x_i \dot{x_j}\right)
         \nonumber\\
        &-& 2\omega^2g_{ij}\left(y_i y_j - y_i x_j  + \frac{1}{2}x_i x_j\right)
        \label{sn5}
\end{eqnarray}
Since there is no kinetic term for $y_i$ it is really an auxiliary variable.
Eliminating $y_i$ from $L(y,x)$ by using its equation of motion
we directly arrive at (\ref{11}).
The soldering formalism automatically ensures that the sum of the hamiltonians corresponding to the elementary modes (\ref{23}) is equal to the hamiltonian of the   composite system. This will be explicit in the following analysis.

The lagrangians of the elementary doublet (\ref{23}) merit careful study. We take $L_+$ for instance.
There are two coordinates $x_1 $and $x_2$. The corresponding momenta are defined as $\pi_1 = \frac{ \partial{L_+}}{\partial{\dot{x_1}}}= -i\omega x_2$ and $\pi_2 =\frac{ \partial{L_+}}{\partial{\dot{x_2}}}= i\omega x_1$. These immediately lead to the two primary constraints
\begin{eqnarray}
\Phi_1 = \pi_1 + i\omega x_2\approx 0\nonumber\\
\Phi_2 = \pi_2 - i\omega x_1\approx 0\label{cons}
\end{eqnarray}
which are second class. The symplectic structure  is given by the Dirac brackets instead of the Poisson brackets. Since $L_+$ is first order, the Dirac brackets can be read off from the following modified form of $L_+$, which is obtained by discarding a total time derivative,
\begin{equation}
L_+ =2i\omega x_1\dot{x}_2 -\omega^2\left({x_1}^2 - {x_2}^2\right)
\end{equation}
Clearly, the symplectic structure is given by \cite{FJ}
\begin {equation}
\{ x_1, x_2\} =  \frac{i}{2\omega}\label{Bnewm}
\end{equation}
and the hamiltonian 
\begin {equation}
\tilde{{\cal{H}}} = \omega^2(x_1^2 - x_2^2)\label{Hnew}
\end{equation}
From the symplectic structure (\ref{Bnewm}) it is evident that $-2i\omega x_2$ is the conjugate momentum to $x_1$. Now consider the canonical transformation to $\left( x_+, p_+\right)$, where
\begin {equation}
x_+ =  i\sqrt{2} x_2 \hskip .3cm \rm{and}\hskip .3cm p_+ =\sqrt{2}\omega x_1 \label{Bnew2}
\end{equation}
In terms of the new canonical variables, the Hamiltonian (\ref{Hnew}) becomes 
\begin{equation}
{\cal{H_{+}}} = \frac{1}{2}p_{+}^2 +\frac{1}{2}\omega^2 x_{+}^2\label{30new+}
\end{equation}
Note that though the system $L_+$  is a two coordinates system it has one degree of freedom in the configuration space. This can be easily verified from the rule of degrees of freedom counting for the constrained systems. Another remarkable aspect is the symplectic structure (\ref{Bnewm}). Though both $x_1$ and $x_2$ are real their Dirac bracket is complex. This is due to the complex constraint algebra which owes to the fact that the original Lagrangian is complex. 

Similar analysis with $L_-$ reveals thst it is also a system with one degree of freedom which also has the same hamiltonian (\ref{Hnew}) but a different symplectic structure,
\begin {equation}
\{ x_1, x_2\} = - \frac{i}{2\omega}\label{Bnew1}
\end{equation}
By a canonical transformation from $(x_1, 2i\Omega x_2)$ to $(x_, p_)$ where 
\begin {equation}
x_- = - i\sqrt{2} x_2 \hskip .3cm \rm{and}\hskip .3cm p_- =\sqrt{2}\omega x_1 \label{Bnew22}
\end{equation} the Hamiltonian of the system may be cast as 
 \begin{equation}
{\cal{H_{-}}} = \frac{1}{2}p_{-}^2 +\frac{1}{2}\omega^2 x_{-}^2\label{30new-}
\end{equation}

 A remarkable feature of the pieces (\ref{23}) is that, like  the composite Lagrangian (\ref{11}),
they are also invariant under the transformation (\ref{24}). Corresponding conserved charges can easily be found using Noether's theorem. If $L({x_i, \dot{x}_i})$ is the lagrangian and $x_i \to x_i^{\prime} = x_i + \delta x_i$ is a symmetry, then 
\begin{equation}
\frac{d}{dt}\left(\frac{\partial L}{\partial\dot{x}_i}\delta x_i\right) = 0\label{2222}
\end{equation}
which gives the conserved charge.
From the symmetry transformation (\ref{24}) we get $\delta x_i = \theta \sigma_{ij}x_j$. The Noether charges $C_{\pm}$ following from
(\ref{23}) can be easily derived. These may be written as
\begin{equation}
C_{\pm} = \pm\frac{\tilde{{\cal{H}}}}{\omega}\label{222}
\end{equation}
where $\tilde{{\cal{H}}}$  is the 
Hamiltonian following from
${\cal{L_{\pm}}}$ (see equation(\ref{Hnew})).
  The doublets (\ref{23}) have opposite 'charges' w.r.t. the
SU(1,1) transformations in the plane. They may be aptly called as
 the pseudo - chiral oscillators.
The pseudo - chiral oscillators thus manifest dual aspects of the
SU(1,1) symmetry of (\ref{11}).
This opposite (pseudo) chirality of the elementary modes will play a
 crucial role in our construction of the dynamical realization of the J--S algebra.  
{\vspace {0.7cm}}

 From the above we have seen that the Lagrangian of the bidimensional
oscillator in the indirect approach, writtwn in terms of the  hyperbolic
coordinates, can be viewed as a 
 coupling of the independent pseudo-chiral doublet
(\ref{23}). We will now show that the same reduction holds from the hamiltonian point of view. The hamiltonian analysis derives its significance from the necessity of showing the internal consistency of the soldering formalism. More important is the fact that 
the hamiltonian analysis is the precursor of quantization.

 We start from the lagrangian (\ref{11}). The  Hamiltonian
obtained from (\ref{11}) is given by 
\begin{eqnarray}
H_I &=& \frac{1}{2}g_{ij}\left(p_ip_j +  \omega^2x_ix_j\right)\nonumber\\
    &=& \left(\frac{1}{2}p_1^2 +\frac{1}{2}\omega^2x_1^2\right) 
- \left(\frac{1}{2}p_2^2 + \frac{1}{2}\omega^2x_2^2\right)
\label{27}
\end{eqnarray}
where $p_i$ is the momenta conjugate to $x_i$.
Note that $H_I$ is equivalent to the {\it{difference}}
 of the hamiltnian of two one - dimensional oscillators \cite{BGPV, BM, amz}. 
Making a canonical transformation
\begin{eqnarray}
p_{\pm} =\frac{1}{\sqrt{2}}p_1 \pm i
             \frac{\omega}{\sqrt{2}}x_2\nonumber\\
x_{\pm} =\frac{1}{\sqrt{2}}x_1 \pm i
             \frac{1}{\sqrt{2}}\frac{p_2}{\omega}\label{ct}
\end{eqnarray} 
it is possible to diagonalise $H_I$ as,
\begin{equation}
H_I = {\cal{H_+}} + {\cal{H_-}}\label{29}
\end{equation}
where
\begin{equation}
{\cal{H_{\pm}}} = \frac{1}{2}p_{\pm}^2 +\frac{1}{2}\omega^2 x_{\pm}^2\label{30}
\end{equation}
Clearly, they are the same as those given by (\ref{30new+}) and (\ref{30new-}). The reduction of (\ref{11}) to the doublet (\ref{23}) is thus established from the hamiltonian approach.

Looking back to the reduction process we find that the price one has to pay is that the canonical variables $x_{\pm}$
and $p_{\pm}$ are no longer real. As a result
 the Hamiltonians $ {\cal{H_{\pm}}}$ are not hermitian. Note, however, that
\begin{equation}
{\cal{H_{\pm}}}^{\dagger}= {\cal{H_{\mp}}}\label{31}
\end{equation}
so that the hermiticity of $H_I$ is preserved.
One can further prove that
\begin{equation}
\eta^{-1} {\cal{H_{\pm}}}^{\dagger} \eta
 = {\cal{H_{\pm}}}\label{ph}
\end {equation}
where $\eta = PT$. The above condition is a consequence of the fact that under $PT$ transformation
\begin{equation}
\eta x_i \eta^{-1} =  g_{ij}x_j,\hspace{.3cm} \eta p_i \eta^{-1} = - g_{ij}p_j
\end{equation}
following from (\ref{24d}).
Hence  
\begin{equation}
\eta x_{\pm} \eta^{-1} = x^{\dagger}_{\pm}\hspace{.3cm}{\rm{and}}\hspace{.3cm}
   \eta p_{\pm} \eta^{-1} =  p^{\dagger}_{\pm}\label{pc}
\end{equation}
Basing on (\ref{ph}) it is possible to
build a consistent quantum mechanics \cite{BM}.

To facilitate a better understanding of the quantisation procedure it will be useful to give some details of the quantization formalism that has been applied here \cite{BM}. The states are represented by ket vectors  $\ket{\psi}$ in the usual way. The corresponding bra vectr is obtained by complex conjugation. Both the bra and ket vectors form linear vector spaces i.e.$\ket{\alpha}$ and $\ket{\beta}$ are two ket vectors of V, then
\begin{eqnarray}
c\ket{\alpha} + d\ket{\beta} \epsilon V
\end{eqnarray}
Also,
\begin{eqnarray}
c\bra{\alpha} + d\bra{\beta} \epsilon V_D
\end{eqnarray}
where $V_D$
is dual to $V$ and $c,d$ are complex numbers. Note that c-numbers always multiply the elements of $V$ and $V_D$ from the left. 

 Since we are working with complex operators where $\eta$ hermiticity ia relevant rather than the usual hermiticity, we define the $\eta$ hermitian conjugate of an operator as
 \begin{equation}
\tilde{A} = \eta^{-1} A^{\dagger} \eta\label{phc}
\end{equation}
IF $\tilde{A} = A$, $A$ will be called $\eta$ hermitian.
The rule of
dual correspondence
will be modified from the usual one.
If $\ket{\alpha}$ and $\ket{\beta}$ are two ket vectors, their scalar product will be defined as $\braket{\bar{\beta}}{\alpha}$, where $\ket{\bar{\beta}}$ is defined as the $\eta$
transformed
ket
\begin{equation}
\ket{\bar{\beta}} = \eta\ket{\beta}\label{connection}
\end{equation}
 The definition of dual correspondence follows from this:
\begin{equation}
\ket{\alpha} \Longleftrightarrow   \bra{\bar{\alpha}}
\end{equation}
We will see in the following that this definition is consistent with the observables being associated with $\eta$ hermitian operators i.e operators which satisfy $\tilde{A} = A$. The scalar product thus defined will be assumed to satisfy the usual properties of the scalar product. Specifically, $ \braket{\bar{\alpha}}{\beta}$ = $\braket{\beta}{\bar{\alpha}}^*$. This automatically leads to $\braket{\bar{\alpha}}{\alpha}$ to be real. We further assume $\braket{\bar{\alpha}}{\alpha} \geq 0$.

The operator $\eta$ involves the
time reversal operator. In the usual analysis in quantum mechanics the combination $\eta^{-1}A\eta$ is consistently interpreted when assumed
to act on ket only \cite{SA}. We will follow the same convention here.

To further elucidate the properties of the state space we introduce a basis formed by eigenvectors of a  $\eta$ hermitian operator $\hat{O}$. Since it is a non hermitian operator the basis is bi - dimensional: $\{ \ket{\psi_n}, \ket{\bar{\psi}_n}\}$, where
\begin{eqnarray}
\hat{O}\ket{\psi_n} &=& O_n\ket{{\psi}_n}\nonumber\\
{\hat{O}^{\dagger}}\ket{\bar{\psi}_n} &=& {O_n}^*\ket{{\bar{\psi}}_n}\label{ev}
\end{eqnarray}
Using (\ref{connection}) it is easy to show that (\ref{ev}) is a consequence of $\eta$ hermiticity of $O$. The orthonormality and completeness relations are
\begin{eqnarray}
\braket{\psi_n}{\bar{\psi}_n} &=& \braket{\bar{\psi}_n}{\psi_n} = \delta_{nk}\nonumber\\
\ket{\psi_n}{\bra{\bar{\psi}_n}} &=& \ket{\bar{\psi}_n}\bra{\psi_n} = I\label{completeness}
\end{eqnarray}

Let us first demonstrate the consistency of our definition of scalar product. Expand $\ket{\alpha}$ in the $\ket{\psi}_n$ basis and $\bra{\bar{\beta}}$
in $\bra{\bar{\psi}}_n$ basis
\begin{equation}
\ket{\alpha} =\Sigma c_n\ket{{\psi}_n} ;\hskip .5cm\bra{\bar{\beta}} = \Sigma d_n\bra{{\bar{\psi}}_n}
\end{equation}
Then 
\begin{equation}
\braket{\bar{\beta}}{\alpha} = \Sigma\Sigma d_n\bra{\bar{\psi}_n}c_k\ket{{\psi}_k} = \Sigma_n\Sigma_k d_nc_n^*\braket{{\psi}_n}{\bar{\psi}_k}= \Sigma_n d_n{c_n}^*
\end{equation}
From the last expression it is easy to see that if $\alpha$ = $\beta$ then $d_n = c_n$. Hence
\begin{equation}
\braket{\bar{\alpha}}{\alpha} = \Sigma_n \abs{c_n}^2
\end{equation}
which is positive definite. This shows the internal consistency of our formalism.

We wikk now show that the eigenvalues of the $\eta$ hermitian operators are real, where use will be made of the fact that $\eta = PT$.
Let 
$O$ be such an operator. Then from the second equation of (\ref{ev}) we get 
\begin{equation}
\bra{\bar{\psi}_n} \hat{O} = O_n \bra{\bar{\psi}_n}
\end{equation}
Multiplying the above equation from the right by $\ket {\psi_k}$ we get
\begin{equation} \matrixel{\bar{\psi}_n}{\hat{O}}{\psi_k} ={{O_n}}\delta_{nk}
\end{equation}
while multiplying the first equation the first equation of (\ref{ev}) by $\bra{\bar{\psi}_k}$
from the left we would get
\begin{equation} \matrixel{\bar{\psi}_k}{\hat{O}}{\psi_n} ={{O_n}^*}\delta_{nk}
\end{equation}
where we have used the orthonormality relations given earlier. Compairing the last  two equations we get $O_n = {O_n}^*$.

Now there is an interesting point to observe. Suppose we have a hermitian operator $O_H$ i.e. $O_H^\dagger = O_H$. In our formalism where the dual correspondence is established in an alternative way hermitian operators may not have real value \footnote{The theorem that hermitian operators must have real eigenvalues assumes the usual dual correspondence}. We will show that if a hermitian operator has a real eigenvalue it can not be simultaneously  $\eta$ hermitian. 
 Let $O_H$  have real eigenvalues,
\begin{equation}
O_H\ket{\alpha} = h\ket{\alpha} 
\end{equation} 
where $h$ is a real number. Suppose, if possible, it is simultaneously $\eta$ hermitian. If it is simultaneously $\eta$ hermitian
then $O_H$ commutes with $\eta$. Then $\eta\ket{\alpha}$ belongs to the same eigenvalue as  $\ket{\alpha}$. As the system has one degree of freedom there can not be any degeneracy \cite{LL}. So we conclude
\begin{equation}
\eta\ket{\alpha} = \ket{\alpha}
\end{equation}
But this goes against our definition of dual correspondence. Hence a hermitian operator having  real eigenvalues can not simultaneously be $\eta$ hermitian. A notable example is the identity operator $I$. It is hermitian and has real eigenvalue (which is trivially 1 for all elements of the state space). The operator $\eta^{-1} I \eta$ thus cannot be equated with $I$. One has to treat $\eta^{-1} I \eta$ as a block which has complex eigenvalue.

  The chiral oscillators carry dual aspects of SU(2) and provide 
dynamical structures for the J - S consruction of SU(2) \cite{SK}. Clearly, the
pseudo chiral oscillators (\ref{23}) are candidates for SU(1,1) as
they also serve as the 'spin' doublets with respect to SU(1,1) transformations.
We introduce the operators
\begin{equation}
a = \sqrt{\frac{\omega}{2}}\left(x_+ + \frac{ip_+}{\omega}\right)\label{37}
\end{equation}
and
\begin{equation}
b = \sqrt{\frac{\omega}{2}}\left(x_- + \frac{ip_-}{\omega}\right)\label{38}
\end{equation}
{\footnote{Details of the quantization formalism is given in the appendix}}
The $\eta$ hermitian conjugates of $a$ and $b$ are obtained from the
definition (\ref{phc}) as 
$\tilde{a}$ and $\tilde{b}$ respectively where 
\begin{equation}
\tilde{a} = \eta ^{-1} a^{\dagger} \eta\label{pha}
\end{equation}
\begin{equation}
\tilde{b} = \eta ^{-1} b^{\dagger} \eta\label{pha1}
\end{equation}
From (\ref{37}) and (\ref{38}) we get on using (\ref{pc})
\begin{equation}
\tilde{a} = \sqrt{\frac{\omega}{2}}\left(x_+ - \frac{ip_+}{\omega}\right)
                                 \label{371}
\end{equation}
and
\begin{equation}
\tilde{b} = \sqrt{\frac{\omega}{2}}\left(x_- - \frac{ip_-}{\omega}\right)
                                  \label{381}
\end{equation}
It is easy to prove the algebra
\begin{equation}
[a,\tilde{a}] = [b,\tilde{b}] = 1\label{33n}
\end{equation}
with all other brackets being zero.
The algebra between the operators $a,\tilde{a}, b, \tilde{b} $ can be utilised to construct the operators $N_+ = \tilde{a} a$ and $N_- = \tilde{b} b$, the eigenvalues of which can be shown to be non negetive integers $n_\pm$ \cite{BM}. Thus $N_\pm$ may be interpreted as the number operator for the pseudo - chiral doublet (\ref{23}). 
Also, the hamiltonians can be diagonalised in terms of the number operators as 
\begin{eqnarray}
{\cal{H_+}} &=& \omega \left( \tilde{a} a + \frac{1}{2}\right)\nonumber\\
{\cal{H_-}} &=& \omega \left( \tilde{b} b + \frac{1}{2}\right)
\label{hp}
\end{eqnarray}
From this 
the energy eigenvalue spectra of ${\cal{H}_\pm}$ can be worked out as
\begin{equation}
 E_\pm =\omega \left(n_\pm + \frac{1}{2}\right)
\label{hm}
\end{equation}
where $n_\pm$ are the eigenvalues of $N_\pm$.
Henceforth, as is often done, we will subtract out the zero point energy.

We have now set the stage to construct the representations sought for.
Let us define the following generators
\begin{eqnarray}
J_z &=& \frac{1}{2}\left(\tilde{a}a - \tilde{b}b\right)\nonumber\\
J_+ &=& \tilde{a} b\nonumber\\
J_- &=& -\tilde{b}a \label{40}
\end{eqnarray}
where,
\begin{equation}
J_{\pm} = J_x \pm i J_y
\label{pm}
\end{equation}
These generators are $\eta$ hermitian, as may be explicitly verified.
Furthermore, these operators satisfy
\begin{equation}
\left[J_z,J_{\pm}\right] = \pm J_{\pm},\hspace{.3cm}\left[J_+,J_-\right]=-2J_z
                    \label{41}
\end{equation}
which is nothing but the SU(1,1) algebra. The construction (\ref{40})
is then a realization of the SU(1,1) algebra based on the algebra (\ref{33n}).

At this point one should observe that the realization (\ref{40}) is a dynamical realization of the algebra. The operators $(a, \tilde{a})$ and ($b,\tilde{b}$) are
respectively the annihilation and creation operators belonging to the
plus (minus) type pseudo - chiral oscillators (\ref{23}). This identification can be substantiated in the following way. From (\ref{222}) we see that the dynamical model predicts opposite spin projections. The corresponding 
Hamiltonians have been shown to be 
${\cal{H_{\pm}}}$ given by 
(\ref{30}). These Hamiltonians
have been diagonalized above (see equations (\ref{hp}) in terms of the operators $a$ and $b$ and their eigenvalue spectrum has been worked out in (\ref{hm}). On the other hand the first equation of the set (\ref{40}) shows on the algebraic level that the two types of excitations carry opposite spin projections. 

 The dynamical basis of our construction is further butressed from the structure of the Casimir operator of the algebra. We can show that the operator $J^2$ defined by
\begin{eqnarray}
J^2 &=& J_z^2 - \frac{1}{2}( J_+J_- + J_-J_+)\nonumber\\
    &=& \left[\frac{1}{2}\left(\tilde{a}a + \tilde{b}b\right)\right]
     \left[\frac{1}{2}\left(\tilde{a}a + \tilde{b}b\right) + 1 \right]
       \label{c2}
\end{eqnarray}
commutes with all three generators $J_x, J_y $ and $J_z$. In other words $J^2$ is the Casimir operator of SU(1,1). Note that $J^2$ is factorised as
\begin{equation}
J^2 = \frac{N}{2}
     \left(\frac{N}{2} + 1\right),\hskip .3cm N =N_+ + N_- 
                    \label{41new}
\end{equation}
Comparing the above with the well known form of the
Casimir operator of SU(2) 
, we can recognize the 
 structural similarity between the expressions of the Casimir operators
in terms of the basic variables 
for SU(2) \cite{SK} and (\ref{c2}). Indeed, for SU(2) the Casimir, in terms of the usual oscillator variables, is given by
\begin{eqnarray}
J^2 &=& \left[\frac{1}{2}\left({a}^{\dagger}a + {b}^{\dagger}b\right)\right]
     \left[\frac{1}{2}\left({a}^{\dagger}a + {b}^{\dagger}b\right) + 1 \right]
       \nonumber\\
      &=&  \frac{N}{2}
     \left(\frac{N}{2} + 1\right)
\end{eqnarray}
Here, $\left(a, a^\dagger\right)$ and $\left(b, b^\dagger\right)$ are the two independent pairs of creation/annihilation operators. Their analogues are $\left(a, \tilde{a}\right)$ and $\left(b, \tilde{b}\right)$ for SU(1,1), as already pointed out.
This reveals again the exact parallel
between our constructions of SU(2) and SU(1,1) algebras based on the
dynamical structures of the chiral or pseudo - chiral oscillators.

 Using (\ref{pm}) we can explicitly determine $J_x$, $J_y$ and $J_z$
from (\ref{40}) as
\begin{eqnarray}
J_x &=& \frac{1}{2}\left(\tilde{a}b - \tilde{b}a\right)\nonumber\\
J_y &=& -\frac{i}{2}\left(\tilde{a}b + \tilde{b}a\right)\nonumber\\
J_z &=& \frac{1}{2}\left(\tilde{a}a - \tilde{b}b\right)\label{x}
\end{eqnarray}
From the expressions (\ref{37}), (\ref{38}),(\ref{371}) and (\ref{381})
we find that $J_x$, $J_y$ and $J_z$ are, expectedly, $\eta$  hermitian.
Interestingly, it is possible to construct, from the above, a realization of the SU(1,1) algebra 
consisting of hermitian operators only, by the following 
mapping
\begin{eqnarray}
J_x &\to & J_y\nonumber\\
J_y &\to & iJ_z\nonumber\\
J_z &\to & -iJ_x\label{map2}
\end{eqnarray}
Explicitly, the new hermitian generators are,
\begin{eqnarray}
J_x &=& \frac{i}{2}\left(\tilde{a}a - \tilde{b}b\right)\nonumber\\
J_y &=& \frac{1}{2}\left(\tilde{a}b - \tilde{b}a\right)\nonumber\\
J_z &=& -\frac{1}{2}\left(\tilde{a}b + \tilde{b}a\right)\label{x1}
\end{eqnarray}
 That the mapping (\ref{map2}) preserves the SU(1,1)
algebra can be seen from (\ref{41}). Alternatively, one can check it
directly from (\ref{x1}).
One may now verify that with this new hermitian representation the factorisability property of the Casimir operator (\ref{c2}) still holds. This shows the robustness of our scheme.

   At this point it is instructive to compare our representation (\ref{40})
with the usual J - S realization of SU(1,1). The latter is given by \cite{SBJ}
\begin{eqnarray}
J_z &=& \frac{1}{2}\left(a^{\dagger}a + bb^{\dagger}\right)\nonumber\\
J_+ &=& J_x + iJ_y = a^{\dagger}b^{\dagger} \nonumber\\
J_- &=& J_x - iJ_y = a b\label{340}
\end{eqnarray}
This representation is based on two independent harmonic oscillator algebras.
Note that, in contrast to (\ref{40}), (\ref{340}) 
cannot be interpreted in terms of independent
dynamical structures. This can be seen very simply by writing the Casimir
operator
from (\ref{340})
\begin{eqnarray}
C &=& J_z^2 - \frac{1}{2}\left(J_+J_- + J_-J_+\right)\nonumber\\
  &=& \frac{1}{4}\left(a^{\dagger}a - b^{\dagger}b\right)^2 - 
          \frac{1}{2}\left(a^{\dagger}a + b^{\dagger}b + 1\right)\label{c3}
\end{eqnarray}
Clearly this cannot be factorised as (\ref{c2}). On the other hand, the 
Casimir operator
 obtained from our realization 
(\ref{40}) (or (\ref{x1})) factorises properly.
 Our realizations are
therefore fundamentally different from the usual one ( equation (\ref{340})). 

\bigskip

 Our paper provides an instructive example of how modern notions of $PT$ symmetry and pseudo hermiticity play an important role leading to fresh insights and new results even in familiar systems that follow standard quantization prescriptions. Taking the harmonic oscillator as an example we show that its quantization based on $PT$ symmetry can be exploited to construct a new Jordan - Schwinger (J - S) map of SU(1,1) algebra. This is shown to be fundamentally different from the usual J - S map for SU(1,1).

 The Jordan - Schwinger realisation \cite{js1, js2} of the SU(1,1) group presented here is endowed with a dynamical structure behind it which consists of a doublet of one dimensional pseudo - chiral oscillators. These components together form the indirect representation of a bidimensional oscillaor. The system is shown to have SU(1,1) invariance, a result which is much less emphasised in the literature than its SU(2) invariance. We reduce the lagrangian of the planar oscillator in its elementary modes which carry opposite aspects of the SU(1,1) rotation symmetry. These elementary modes constitute the basic building blocks of the dynamical realisation
of the SU(1,1) algebra.

   The hamiltonians corresponding to the elementary lagrangians are characterised by a remarkable feature -- they are non-hermitian but are $PT$ symmetric. In the recent times much work has been done in the study of quantum mechanics with non-hermitian hamiltonians which have $PT$ symmetry \cite{B}. Such hamiltonians have been called $\eta$ hermitian here.
    The complex hamiltonians of the elementary modes have been demonstrated to be $PT$ symmetric. We have utilised a quantization procedure introduced by us \cite{BM} in the recent past  which is based on $\eta$ hermiticity. For easy reference an outline of the quantization method is given. In this framework an operator formulation of the harmonic oscillator has been developed which mimics the Dirac-Heisenberg quantization of the harmonic oscilator. Operators acting as the raising and lowering operators have been invoked and their algebra worked out. The new Jordan - Schwinger realisation of the SU(1,1) group is then provided where the generators are constructed out of a doublet of such oscillator algebras.

Our work gives a holistic picture of constructing the J - S representations from the bidimensional oscillator. The well known representation of SU(2) is realised in the realm of familiar quantum mechanics \cite{js1, js2, SK}. On the other hand quantization of the same  bidimensional oscillator in the framework of $\eta$ hermitian hamiltonian leads to the J - S representation of SU(1,1).
In both the cases factorisability of the Casimir comes out as expected from general arguments.

\vspace {.7cm}

 

\end{document}